\documentclass[aps,twocolumn,twoside,nofootinbib,prl,preprintnumbers]{revtex4}

\newcommand{\lsim}{\lesssim}
\newcommand{\gsim}{\gtrsim}

\newcommand{\Mpl}{M_{\rm Pl}}

\newcommand{\ra}{\rightarrow}
\newcommand{\Ra}{\Rightarrow}

\newcommand{\three}{\mathbf{3}}
\newcommand{\threebar}{\overline{\mathbf{3}}}

\begin{document}
\bibliographystyle{revtex}

\preprint{hep-ph/0304256}

\title{Rationalizing right-handed neutrinos}

\author{Graham D. Kribs}
\affiliation{\mbox{Department of Physics, University of Wisconsin, 
Madison, WI 53706} \\ 
\mbox{\textnormal{\texttt{kribs@physics.wisc.edu}}}}


\begin{abstract}
A simple argument based on an SU(3) gauged horizontal symmetry 
is presented that connects the explanation for three generations 
of matter with the existence of a triplet of right-handed neutrinos.
This rationale for right-handed neutrinos is analogous to, 
but completely independent of, grand unification or extra universal 
dimensions.  A brief discussion of the supersymmetrized SU(3) model 
is also given, pointing out that certain problems in ordinary 
supersymmetric models such as fast proton decay via dimension-5 
Planck-suppressed operators can be naturally solved.

\end{abstract}


\maketitle


One of the most remarkable features of the Standard Model (SM)
is that matter fermions are chiral and yet all gauge 
\cite{gauge,witten} and gravitational \cite{aw} anomalies 
vanish for each generation.  A known but not often 
emphasized fact about the matter content is that, given
one generation with unfixed hypercharges, anomaly cancellation 
determines the relative hypercharge assignment to be precisely
what has been established by experiment \cite{SManomalies}.  
In other words, electric charge quantization is essentially 
automatic without grand unification.  
This fact, taken at face value, is circumstantial evidence
against the existence of right-handed neutrinos.  By definition a 
candidate for a right-handed neutrino is any fermion that is uncharged 
under all of the SM gauge symmetries.  Yet, gauge symmetries are
precisely the reason that each type of matter ($Q$, $u$, $d$, $L$, $e$) 
is tied with the other matter fields together in a self-consistent, 
exclusive fashion.  In addition, non-chiral matter allows a new mass 
scale unconnected to electroweak symmetry breaking that only further 
complicates our understanding of mass generation and mass hierarchies.
Extensions of the SM with non-chiral matter, such as adding right-handed 
neutrinos, therefore appear to be contrary to all of the guiding wisdom  
gleaned from experiment, at least until recently.
(Those who are still in doubt need only observe the agony that 
the $\mu$ problem causes avatars of supersymmetry.)

Neutrino experiments \cite{superk,sno,kamland}, 
however, have firmly established that 
the neutrinos oscillate between each generation and thus they
have mass.  The largest mass of any one neutrino is constrained 
to be less than about 2 eV \cite{mass}, and more likely their mass is 
one to a few orders of magnitude below this, depending on the generation.
The mechanism of mass generation for neutrinos remains a mystery.
If neutrinos acquire mass analogously to the SM matter fermions, preserving
lepton number, then the particle content
must be extended with at least two right-handed neutrinos $\nu_{1,2}$.
Ordinary Yukawa terms ${\cal L} = \lambda_\nu L H \nu^c$ with tiny 
couplings $\lambda_\nu \lsim 10^{-12}$ suffice to explain the two 
undisputed mass differences found in neutrino oscillation experiments.  

But, the global symmetry behind lepton number conservation is not 
expected to be exact.
At dimension-5, the operator $H L H L/M$ violates lepton number 
by two units and leads to a tiny Majorana mass $v^2/M$ for 
left-handed neutrinos.  This transmutes the neutrino mass 
hierarchy problem from explaining $\lambda_\nu \lsim 10^{-12}$ to 
instead explaining $v/M \lsim 10^{-12}$.
To embrace the dimension-5 neutrino mass explanation means the SM 
effective theory breaks down at $M \lsim 10^{14}$ GeV\@.  This is 
somewhat disconcerting since there are dimension-6 operators that 
violate lepton and baryon number, leading to a proton decay 
rate that is excluded by experiment unless $M \gsim 10^{16}$ GeV\@.
Hence, while lepton number must be be violated at $M$ to explain
neutrino masses, baryon number must be preserved to 
keep the proton stable.  The simplest phenomenological explanation for 
lepton number violation without baryon number violation at the cutoff scale 
$M$ is to add right-handed neutrinos to the SM with ordinary Yukawa couplings, 
plus a heavy Majorana mass term ${\cal L} = M \nu^c \nu$ 
for the right-handed neutrinos.
The resulting combination of a Dirac mass and a heavy Majorana 
mass leads to the famous ``see-saw'' neutrino mass matrix \cite{seesaw}
\begin{equation}
\left( \begin{array}{cc} 0 & \lambda_\nu v \\ \lambda_\nu v & M \end{array} 
\right) \; .
\end{equation}
Diagonalizing this mass matrix, or equivalently integrating out the 
right-handed neutrino gives back the SM plus the dimension-5 operator 
$H L H L/M$ with a well-defined coefficient, $\lambda_\nu^2$, and thus 
predicted neutrino mass, $\lambda_\nu^2 v^2/M$.
A right-handed neutrino with Majorana mass $M$ therefore provides an 
ultraviolet completion of the effective theory beyond the cutoff $M$, 
explaining why only lepton number was violated at $M$.

The difficulty with such neutrino mass generation mechanisms is that
they do not really solve the neutrino mass hierarchy problem, and 
worse still, require precisely those odd-ball fields -- right-handed 
neutrinos -- that are unconnected to SM matter through gauge anomalies.  
Furthermore, the see-saw explanation requires a new Majorana mass scale 
unconnected with electroweak symmetry breaking.
These facts would ordinarily be highly 
distressing except for a remarkable coincidence:  
the scale $M \sim 10^{14}$ GeV is tantalizingly close to where 
the SM gauge couplings come to an approximate intersection.
Such an intersection is predicted by grand unified theories (GUTs), 
providing justification for the new scale.  Furthermore, in an SO(10) GUT 
each right-handed neutrino is elegantly fused with each generation 
of SM matter into a single ${\bf 16}$ representation \cite{so10}.  
This is really just 
an artifact of unifying into a GUT group with rank greater than that 
of the SM, since candidates for right-handed neutrinos in GUTs are 
those fields uncharged under the SM symmetries but charged under some
additional gauge symmetry [for SO(10) this is the extra U(1) under 
the decomposition SO(10) $\ra$ SU(5) $\times$ U(1)].
Rank $>4$ GUTs therefore provide a rationale for $n = 0$ mod $3$ 
right-handed neutrinos whenever each generation is unified into 
a single representation of the group.  

Unfortunately, grand unification has many well-known problems of 
implementation.
Non-supersymmetric grand unification proposals suffer from the 
hierarchy problem as well as a rather inexact unification of gauge 
couplings.  Both non-supersymmetric and supersymmetric unification
models predict proton decay at a rate that has been experimentally 
ruled out in the simplest models.  Also, several theoretical 
problems pervade unification ranging from understanding
how the Higgs is embedded into a GUT representation (the doublet-triplet
splitting problem), how (or if) Yukawa couplings are unified, etc.
Such experimental and theoretical problems ought to induce us to reconsider
GUTs as the origin of right-handed neutrinos and the Majorana mass scale.
Is there any rationale \emph{independent} of unification that 
predicts right-handed neutrinos as well as the Majorana mass scale?

Suppose the explanation for the number of generations is that each 
field's three generations ($Q_{1,2,3}$, $u_{1,2,3}$, etc.) correspond 
to three components of a multiplet of a ``horizontal'' flavor symmetry.  
There are only two continuous symmetries that
are suitable for this purpose possessing a $\three$ representation:  
SU(3) \cite{y} and SU(2) $\sim$ SO(3) \cite{wz,ty}.  SU(2) can be 
summarily dismissed if right-handed neutrinos are required to be in 
a chiral representation of the new symmetry.  It has already been 
emphasized that non-chiral fermions, and right-handed neutrinos 
in particular, seem to have no (aesthetic) place in the SM 
if anomaly cancellation is to connect all matter together.  
There is no hope with SU(2) since it is anomaly-free.  SU(2) also does 
not predict the number of generations since representations of any 
dimension are possible \cite{so3comment}.
Instead, SU(3) admits chiral fermions with only certain dimensionality -- 
there can be three but not two, four, five, seven, 
etc.\ generations.  Moreover, SU(3) provides two additional
key ingredients:  
(1) there is an additional anomaly cancellation condition on the 
matter content if SU(3) is at least weakly gauged, and 
(2) all fermion masses, including right-handed neutrino masses, 
arise from spontaneous symmetry breaking.  Before proceeding, note 
that the connection between SU(3) anomaly cancellation and 
the existence of right-handed neutrinos was made some time ago in \cite{y}.
In this paper the argument in presented in detail, contrasting with 
grand unification and universal extra dimensions, and then implications 
for a supersymmetrized version are briefly discussed.

Gauging a new symmetry in which SM fermions transform is non-trivial 
and requires the cancellation of all gauge anomalies associated with 
the new symmetry.  There are potentially eight new gauge 
anomalies associated with SU(3)$_f$:
[SU(3)$_f$]$^3$; 
SU(3)$_f$ $\times$ [SU(3)$_c$]$^2$;
[SU(3)$_f$]$^2$ $\times$ SU(3)$_c$; 
SU(3)$_f$ $\times$ [SU(2)$_L$]$^2$; 
[SU(3)$_f$]$^2$ $\times$ SU(2)$_L$; 
SU(3)$_f$ $\times$ [U(1)$_Y$]$^2$; 
[SU(3)$_f$]$^2$ $\times$ U(1)$_Y$; 
and
SU(3)$_f$ $\times$ [grav]$^2$.
Six of these are trivially zero since $\mathrm{tr}[t^a] = 0$
for SU(N) gauge groups with $N>1$.  This leaves the mixed
flavor symmetry/hypercharge anomaly [SU(3)$_f$]$^2$ $\times$ U(1)$_Y$,
and the [SU(3)$_f$]$^3$ anomaly.  The mixed anomaly leads to a 
condition on the sum of the hypercharges of the SM fermions that 
is equivalent to the mixed [grav]$^2$ $\times$ U(1)$_Y$ anomaly
\begin{eqnarray*}
[SU(3)_f]^2 \times U(1)_Y: & & \!\!\!\!
\left( 6 Y_Q + 3 Y_u + 3 Y_d + 2 Y_L + Y_e \right) = 0 , 
\end{eqnarray*}
and so automatically cancels.
The [SU(3)$_f$]$^3$ anomaly, however, does not cancel with
just the SM fermion content \cite{wz,y}.

This is straightforward to see:
Five types of matter ($Q$, $u$, $d$, $L$, $e$) can be assigned to 
either $\three$ or $\threebar$ representations.  
Two of the five fields contribute an even number of $\three$'s or 
$\threebar$'s to the anomaly ($Q: \pm 6$; $L: \pm 2$) while the 
remaining three fields contribute an odd number of $\three$'s or 
$\threebar$'s ($u: \pm 3$, $d: \pm 3$, $e: \pm 1$).  The sum of two
even numbers and three odd numbers is an odd number, and so
[SU(3)$_f$]$^3$ anomaly $a \propto n_{\three} - n_{\threebar}$
cannot be canceled no matter how SM matter
is assigned to SU(3)$_f$.

The simplest assignment of matter in $\three$ or $\threebar$
representations allows ordinary Yukawa couplings.
Here the Higgs scalar doublet is assumed to be a singlet under 
SU(3)$_f$, since there is no need for (and various reasons that disfavor)
more than one Higgs doublet in the SM\@.  Gauge invariance of the three 
Yukawa couplings of the SM implies three relations among the anomaly 
coefficients of SM matter
\begin{eqnarray}
Q H u^c   &\Ra& a(Q) + a(u^c) = 0 \\
Q H^* d^c &\Ra& a(Q) + a(d^c) = 0 \\
L H^* e^c &\Ra& a(L) + a(e^c) = 0 \; .
\end{eqnarray}
Without loss of generality $Q$ can be chosen to be a $\three$, 
then $u^c$ and $d^c$ must both be $\threebar$'s.  There are
two choices for the leptons:
[$L(\three)$, $e^c(\threebar)$] or [$L(\threebar)$, $e^c(\three)$].
In either case, the [SU(3)$_f$]$^3$ anomaly coefficient becomes
\begin{equation}
a = 6 - 3 - 3 \pm (2 - 1) = \pm 1 \; .
\end{equation}
Notice that the anomaly associated with colored fermions
self-cancels, but with the leptons it does not cancel regardless of
assigning $(L,e^c)$ into a $(\three,\threebar)$ or $(\threebar,\three)$.

Intriguingly, the [SU(3)$_f$]$^3$ anomaly is canceled 
by adding a single new field that transforms as a 
$\threebar$ [for $L(\three)$, $e^c(\threebar)$] or 
$\three$ [for $L(\threebar)$, $e^c(\three)$]  
under SU(3)$_f$. 
To avoid spoiling the SM anomaly cancellation conditions
this field must be neutral under SM gauge symmetries.
Hence, this anomaly-cancellation field has precisely the 
quantum numbers of a right-handed neutrino.
Also, a Yukawa interaction connecting the left-handed with 
the right-handed neutrino, ${\cal L} = L H \nu^c$,
is automatically allowed by SU(3)$_f$ gauge invariance regardless
of the initial choice of 
($\three$, $\threebar$, $\threebar$) versus 
($\threebar$, $\three$, $\three$)
for ($L$, $e^c$, $\nu^c$).

This is a remarkable result.  Let me restate the assumptions and 
the implication:
Assuming
(1) the explanation for the number of generations is a gauged SU(3)$_f$ 
    flavor symmetry, 
(2) all matter is assigned to chiral representations 
    ($\three$ or $\threebar$) of SU(3)$_f$, and 
(3) ordinary Yukawa couplings are SU(3)$_f$ gauge invariant,
then there \emph{must} exist one set of right-handed neutrinos 
$\nu_{1,2,3}$ transforming as a triplet of SU(3)$_f$.

It is important to emphasize that this flavor symmetry rationale 
for right-handed neutrinos is completely independent of grand unification.  
In fact, the simplest assignment that allows Yukawa couplings 
to be gauge invariant under SU(3)$_f$ does not commute with the
usual matter embeddings in unified representations of GUTs. 
For example, SU(5) [as well as SO(10) and $E_6$] unifies $Q$ and $u$ 
into a single representation; this is inconsistent with the SU(3)$_f$ 
assignment given above.  However, Yukawa couplings are notoriously
over-constrained in GUTs as well as flavor symmetry models.  
SU(5) predicts the down and lepton Yukawas of each generation should 
unify, and SO(10) predicts up, down, and lepton Yukawas to unify.  
These predictions are badly broken at low energies, and not much better 
at the GUT scale for all but perhaps $\lambda_b$ and $\lambda_\tau$.
Analogously, the simplest SU(3)$_f$ assignment allows Yukawa couplings 
for all generations, but no generational differences.  This must come from 
additional structure related to the flavor symmetry breaking that has 
not been specified here.  
Nevertheless, the matter (and Higgs) assignments under SU(3)$_f$ can 
be suitably modified to commute with grand unification.
This was done in several early works on gauged SU(3)$_f$ $\times$ 
SU(5) \cite{SU3Hearly}.  There they found that many more 
triplets (or perhaps larger representations) of right-handed neutrinos 
were needed to cancel the [SU(3)$_f$]$^3$ anomaly.
For the purposes of this paper, it is enough to observe that there
must be \emph{at least one} triplet of right-handed neutrinos
to gauge SU(3)$_f$.  

The absence of a signal for new physics in flavor-changing neutral 
current processes places an important constraint on the scale of SU(3)$_f$
symmetry breaking.  The constraint arises from the tree-level
exchange of flavor gauge bosons that lead to transitions between
same-charge, different-generation quarks or leptons.  Integrating
out heavy flavor gauge bosons results in a low-energy effective
theory with new contributions to four-fermion, flavor-violating operators
\begin{equation}
\frac{g_f^4}{M_f^2} (\overline{f}_i \gamma_\mu f_i) 
(\overline{f}_j \gamma^\mu f_j) \; ,
\end{equation}
where $g_f$ is the SU(3)$_f$ gauge coupling and $M_f$ is the symmetry
breaking scale.  If the couplings are CP-conserving, one of the strongest 
constraints comes from the $\Delta s = 2$ process that contributes to the
$K_0-\overline{K}_0$ mass difference.  Estimates of the bound on the 
four-quark operator suggest $M_f \gsim g_f^2 \times 1600$ TeV \cite{lns}.  
The bound is significantly stronger if the couplings maximally violate CP.
In any case, for a flavor gauge coupling that is of order the SM gauge 
couplings, the bound on the symmetry breaking scale is at least 
hundreds of TeV\@.  This is reminiscent of the constraints on extended 
technicolor \cite{de}.

The benefit of right-handed neutrinos transforming under a chiral 
representation of the flavor symmetry is that the Majorana mass  
scale is no longer arbitrary.  The right-handed neutrino Majorana mass 
is generated through
flavor symmetry breaking, analogous to SM fermion masses generated 
through electroweak symmetry breaking.  The scale $M_f$ is not predicted, 
but obviously there is no conflict between the lower bound 
$M_f \gsim 1000$ TeV from flavor-changing constraints and 
the upper bound $M_f \lsim 10^{14}$ GeV needed for a successful see-saw 
explanation of neutrino masses.
If $M_f$ were near the lower bound, future experiments 
could search for deviations from (or as-yet unobserved) flavor-changing 
neutral current processes as a signal for SU(3)$_f$.  This would
require neutrino Yukawa couplings $\lambda_\nu \sim 10^{-4}$ nearer 
in value to their lepton cousins.

How are right-handed neutrino Majorana masses generated from 
flavor symmetry breaking?  Consider a pair of complex scalar fields 
in the fundamental representation $\Sigma_{1,2}(\three)$ that acquire 
unaligned vacuum expectation values.  
This is sufficient to break SU(3)$_f$ $\ra$ nothing.  
A right-handed neutrino mass arises from the dimension-4 operator 
$\epsilon_{ijk} \nu^c_i \nu_j \Sigma_k^*$ replacing $\Sigma$ by its vev.  
Curiously, this two-field breaking model 
gives mass to just two off-diagonal components of the $3 \times 3$ 
Majorana mass matrix in flavor space
\begin{equation}
\left( 
\begin{array}{ccc} & \langle \Sigma_1 \rangle & \langle \Sigma_2 \rangle \\
                   \langle \Sigma_1 \rangle & & \\
                   \langle \Sigma_2 \rangle & & 
\end{array} \right)
\end{equation}
due to the anti-symmetric contraction of SU(3)$_f$ indices.  This may
be a useful starting point for generating an interesting neutrino
mass texture.  Also, the flavor symmetry could be broken in
stages, such as SU(3) $\ra$ SU(2) $\ra$ nothing, that may be
similarly useful for quark, lepton, or neutrino mass textures.

SU(3)$_f$ is not the only rationale for three generations
and three right-handed neutrinos.  In a recent proposal called ``universal 
extra dimensions'' (UED) \cite{univ}, all matter, Higgs, and gauge bosons 
are promoted to six dimensional fields, and the more complicated 
gauge and gravitational anomaly structure of six dimensional theories 
is used to constrain the matter content \cite{dp}.  Ref.~\cite{dp} found 
that cancellation of the global gauge anomaly \cite{en} required the 
number of generations to be $n_g = 0$ mod $3$, 
and cancellation of the pure gravitational anomaly required $n = n_g$ 
fermionic fields uncharged under the SM gauge group.  
This is intriguingly similar to the SU(3)$_f$ symmetry argument, 
since the matter content is similarly restricted by anomaly cancellation 
of a larger symmetry structure.  Other similarities are remarkable:
\cite{dp} required that all matter was chiral in 6-D, analogous to
requiring all matter to be in chiral representations of SU(3)$_f$. 
This led to two possible chirality assignments in UED that are precisely 
analogous to the $\three$ versus $\threebar$ ``chirality'' possibilities 
for the SM fermions under SU(3)$_f$.  Specifically, the quark doublet 
($Q$) must have the opposite chirality to the quark singlets ($u$, $d$), 
and the lepton doublet ($L$) must have the opposite chirality to the 
lepton singlets ($e$, $\nu$).  In UED the lepton doublet could have the 
same or the opposite chirality of the quark doublet, just as here the 
lepton doublet 
could be assigned to the same ($\three$) or opposite ($\threebar$)
representation of the quark doublet.  Finally, the UED rationale for three 
generations and three right-handed neutrinos does not depend on the 
compactification scale, just as the SU(3)$_f$ argument does not depend 
on the flavor symmetry breaking scale.

There are a few important differences between the six dimensional UED model 
and the SU(3)$_f$ model.  The higher dimensional 
nature of UED implies there is an effective theory cutoff scale 
that is only an order of magnitude above
the compactification scale; in the SU(3)$_f$ model, there is no
such restriction.  Several gauge anomalies, such as [SU(2)$_L$]$^2$ U(1)$_Y$  
that are automatically canceled in the SU(3)$_f$ model, are canceled 
in UED only via the Green-Schwarz mechanism with additional matter.  
Finally, the prediction of three generations is not easily extended 
to a supersymmetric six-dimensional ``universal'' model for a variety 
of reasons \cite{dp}, whereas the SU(3)$_f$ model can be quite simply 
supersymmetrized as will be briefly sketched below.

Everything that has been said for the SM with a gauged SU(3)$_f$ 
flavor symmetry also applies to a straightforward extension of the
minimal supersymmetric standard model (MSSM).
This means promoting matter supermultiplets to (anti-)fundamental 
representations of SU(3)$_f$ while the Higgs supermultiplets
remain singlets, in exact analogy with the non-supersymmetric case.
(In the following discussion the same notation is used for 
the MSSM chiral superfields as for the SM fermion fields.)
There are, however, new restrictions on the allowed operators in
the supersymmetrized SU(3)$_f$ model.
The most interesting, model-independent
restriction is that the dimension-5 operators leading to proton decay
\begin{equation}
\frac{Q Q Q L}{\Mpl} \quad , \quad \frac{u^c u^c d^c e^c}{\Mpl} 
\label{dim-5-eq}
\end{equation}
are forbidden by SU(3)$_f$.
Technically the second operator in Eq.~(\ref{dim-5-eq}) could be 
made gauge-invariant if $u^c$ were assigned the conjugate representation
to that of $d^c$ and $e^c$, but this does not happen for the SU(3)$_f$ model
nor for the embeddings of matter into SU(5) or SO(10) representations.
These operators can be made gauge-invariant by adding a pair of 
SU(3)$_f$ breaking 
superfields $\Sigma(\three)$ and $\overline{\Sigma}(\overline{\three})$, 
whereby Eq.~(\ref{dim-5-eq}) becomes 
\begin{equation}
\frac{Q Q Q L \overline{\Sigma}}{\Mpl^2} \quad , \quad 
\frac{u^c u^c d^c e^c \Sigma}{\Mpl^2} \; .
\end{equation}
Below the SU(3)$_f$ symmetry breaking scale, these dimension-6 operators 
map onto the dimension-5 operators above with tiny coefficients of order
$\langle \Sigma \rangle/\Mpl$.  This is sufficient to cure the
fast proton decay problem that results from the ordinarily unsuppressed 
dimension-5 operators.

A supersymmetrized version of the SU(3)$_f$ model has even more
interesting constraints.  All dimension $\le 4$ lepton number violating 
superpotential terms $Q L d^c$, $L L e^c$, and $L H_u$ are forbidden 
by SU(3)$_f$.  
Again, higher dimension operators with SU(3)$_f$ breaking fields will
reintroduce these terms, but (for the first two) this leads to
significant suppression.  If the flavor symmetry were promoted to 
U(3)$_f$, the dimension-4 baryon number violating term $u^c d^c d^c$ would 
also be forbidden.  An exact flavor symmetry could serve in 
precisely the same role as matter parity on superfields ($R$-parity on 
fields).  Of course the flavor symmetry is broken, and this 
reintroduces these so-called $R$-parity violating operators.  It would be 
interesting to see if $R$-parity could be discarded in favor of a 
spontaneously broken U(3)$_f$ flavor symmetry without sacrificing a 
long-lived proton.

In summary, an extension of the Standard Model with an SU(3)$_f$ gauged 
flavor symmetry is presented that explains why there are
three generations of matter and predicts the existence of three 
right-handed neutrinos.  This argument is independent of
grand unification or extra universal dimensions.  The right-handed 
Majorana mass scale results from spontaneous SU(3)$_f$ symmetry 
breaking.  If the breaking scale is ``low'', less than of order 1000 TeV,
deviations in flavor changing neutral current processes are 
expected due to tree-level flavor gauge boson exchange.
It should be emphasized that such a Majorana mass scale is completely 
consistent with the see-saw explanation for neutrino mass generation 
so long as the Dirac masses of the neutrinos are less than but of order 
the muon mass.  This is a perfectly reasonable possibility given that
SU(3)$_f$ has freed us from thinking only in terms of grand unification.
The supersymmetric extension
including a gauged SU(3)$_f$ is straightforward.  The fast proton
decay problem from dimension-5 Planck-suppressed operators is 
automatically cured, and certain $R$-parity violating couplings
are naturally suppressed.  Combining the SU(3)$_f$ gauged flavor 
symmetry with models that attempt to explain the structure of the 
quark, lepton, or neutrino mass matrices is an extremely interesting 
direction left for future work.

\begin{acknowledgments}

I have benefited from discussions with 
B.~Balantekin, V.~Barger, A.~Nelson, Y.~Nir, G.~Shiu, and L.-T.~Wang.
I also thank the CERN theory group for hospitality where part of 
this work was completed.  This work was supported in part by the 
U.S. Department of Energy under contract DE-FG02-95ER40896.

\end{acknowledgments}


\begin{thebibliography}{99}

\bibitem{gauge}
S.~L.~Adler,
Phys.\ Rev.\  {\bf 177}, 2426 (1969);
J.~S.~Bell and R.~Jackiw,
Nuovo Cim.\ A {\bf 60}, 47 (1969);
W.~A.~Bardeen,
Phys.\ Rev.\  {\bf 184}, 1848 (1969);
C.~Bouchiat, J.~Iliopoulos and P.~Meyer,
Phys.\ Lett.\ B {\bf 38}, 519 (1972);
D.~J.~Gross and R.~Jackiw,
Phys.\ Rev.\ D {\bf 6}, 477 (1972).

\bibitem{witten}
E.~Witten,
Phys.\ Lett.\ B {\bf 117}, 324 (1982).

\bibitem{aw}
L.~Alvarez-Gaume and E.~Witten,
Nucl.\ Phys.\ B {\bf 234}, 269 (1984).

\bibitem{SManomalies}
C.~Q.~Geng and R.~E.~Marshak,
Phys.\ Rev.\ D {\bf 39}, 693 (1989).
J.~A.~Minahan, P.~Ramond and R.~C.~Warner,
Phys.\ Rev.\ D {\bf 41}, 715 (1990).

\bibitem{superk}
Y.~Fukuda {\it et al.}  [Super-Kamiokande Collaboration],
Phys.\ Rev.\ Lett.\  {\bf 81}, 1562 (1998)
[arXiv:hep-ex/9807003].

\bibitem{sno}
Q.~R.~Ahmad {\it et al.}  [SNO Collaboration],
Phys.\ Rev.\ Lett.\  {\bf 89}, 011301 (2002)
[arXiv:nucl-ex/0204008].

\bibitem{kamland}
K.~Eguchi {\it et al.}  [KamLAND Collaboration],
Phys.\ Rev.\ Lett.\  {\bf 90}, 021802 (2003)
[arXiv:hep-ex/0212021].

\bibitem{mass}
For a review, see
C.~Weinheimer,
arXiv:hep-ex/0210050.

\bibitem{seesaw}
M.~Gell-Mann, P.~Ramond and R.~Slansky,
Print-80-0576 (CERN);
T.~Yanagida,
Prog.\ Theor.\ Phys.\  {\bf 64}, 1103 (1980);
R.~N.~Mohapatra and G.~Senjanovic,
Phys.\ Rev.\ Lett.\  {\bf 44}, 912 (1980).

\bibitem{so10}
H.~Georgi, Particles and Fields, Proceedings of the APS Division of
Particles and Fields, ed C. Carlson;
H.~Fritzsch and P.~Minkowski,
Annals Phys.\  {\bf 93}, 193 (1975).

\bibitem{y}
T.~Yanagida,
Phys.\ Rev.\ D {\bf 20}, 2986 (1979);
Prog.\ Theor.\ Phys.\  {\bf 64}, 1103 (1980).

\bibitem{wz}
F.~Wilczek and A.~Zee,
Phys.\ Rev.\ Lett.\  {\bf 42}, 421 (1979).

\bibitem{ty}
T.~Maehara and T.~Yanagida,
Prog.\ Theor.\ Phys.\  {\bf 61}, 1434 (1979).

\bibitem{so3comment}
SO(3) is naively better than SU(2), with only odd dimensional representations,
but SO(3) with just $\three$'s is physically indistinguishable from 
SU(2) with just $\three$'s, and so there is no unambiguous 
explanation for three generations.

\bibitem{SU3Hearly}
See e.g.\
J.~L.~Chkareuli,
JETP Lett.\  {\bf 32}, 671 (1980)
[Pisma Zh.\ Eksp.\ Teor.\ Fiz.\  {\bf 32}, 684 (1980)];
Z.~G.~Berezhiani and J.~L.~Chkareuli,
JETP Lett.\  {\bf 35}, 612 (1982)
[Pisma Zh.\ Eksp.\ Teor.\ Fiz.\  {\bf 35}, 494 (1982)];
K.~Tamvakis and G.~Zoupanos,
Phys.\ Lett.\ B {\bf 126}, 314 (1983).
Z.~G.~Berezhiani,
Phys.\ Lett.\ B {\bf 129}, 99 (1983);
M.~Soldate, M.~H.~Reno and C.~T.~Hill,
Phys.\ Lett.\ B {\bf 179}, 95 (1986).

\bibitem{lns}
See e.g.\
M.~Leurer, Y.~Nir and N.~Seiberg,
Nucl.\ Phys.\ B {\bf 420}, 468 (1994)
[arXiv:hep-ph/9310320].

\bibitem{de}
S.~Dimopoulos and J.~R.~Ellis,
Nucl.\ Phys.\ B {\bf 182}, 505 (1982).

\bibitem{univ}
T.~Appelquist, H.~C.~Cheng and B.~A.~Dobrescu,
Phys.\ Rev.\ D {\bf 64}, 035002 (2001)
[arXiv:hep-ph/0012100].

\bibitem{dp}
B.~A.~Dobrescu and E.~Poppitz,
Phys.\ Rev.\ Lett.\  {\bf 87}, 031801 (2001)
[arXiv:hep-ph/0102010].

\bibitem{en}
S.~Elitzur and V.~P.~Nair,
Nucl.\ Phys.\ B {\bf 243}, 205 (1984).



\end{thebibliography}

\end{document}